\documentclass[reprint, superscriptaddress, showkeys, amsmath, amssymb, aps, prl]{revtex4-1}
\usepackage{graphicx}
\usepackage{dcolumn}
\usepackage{bm}

\begin{document}

\title{Quantum Transport in Ambipolar Few-layer Black Phosphorus}

\author{Gen Long}
\affiliation{Department of Physics and Center for Quantum Materials, the Hong Kong University of Science and Technology, Hong Kong, China}

\author{Denis Maryenko}
\affiliation{Riken Center for Emergent Matter Science (CEMS), Wako 351-0198, Japan}

\author{Sergio Pezzini}
\affiliation{High Field Magnet Laboratory (HFML-EMFL) and Institute for Molecules and Materials, Radboud University, Nijmegen, 6525 ED, the Netherlands}

\author{Shuigang Xu}
\author{Zefei Wu}
\author{Tianyi Han}
\author{Jiangxiazi Lin}
\author{Yuanwei Wang}
\author{Liheng An}
\affiliation{Department of Physics and Center for Quantum Materials, the Hong Kong University of Science and Technology, Hong Kong, China}
\author{Chun Cheng}
\affiliation{Department of Materials Science and Engineering, and Shenzhen Key Laboratory of Nanoimprint
Technology, South University of Science and Technology, Shenzhen 518055, China}
\author{Yuan Cai}
\affiliation{Department of Physics and Center for Quantum Materials, the Hong Kong University of Science and Technology, Hong Kong, China}
\author{Uli Zeitler}
\affiliation{High Field Magnet Laboratory (HFML-EMFL) and Institute for Molecules and Materials, Radboud University, Nijmegen, 6525 ED, the Netherlands}

\author{Ning Wang}
\email[Correspondence to: ]{phwang@ust.hk}
\affiliation{Department of Physics and Center for Quantum Materials, the Hong Kong University of Science and Technology, Hong Kong, China}

\date{\today}

\begin{abstract}
Few-layer black phosphorus possesses unique electronic properties giving rise to distinct quantum phenomena and thus offers a fertile platform to explore the emergent correlation phenomena in low dimensions. A great progress has been demonstrated in improving the quality of  hole-doped few-layer black phosphorus and its quantum transport studies, whereas the same achievements are rather modest for electron-doped few-layer black phosphorus. Here, we report the ambipolar quantum transport in few-layer black phosphorus exhibiting undoubtedly the quantum Hall effect for hole transport and showing clear signatures of the quantum Hall effect for electron transport. By bringing the spin-resolved Landau levels of the electron-doped black phosphorus to the coincidence, we measure the spin susceptibility $\chi_s=m^\ast g^\ast=1.1\pm0.03$. This value is larger than for hole-doped black phosphorus and illustrates an energetically equidistant arrangement of spin-resolved Landau levels. Evidently, the n-type black phosphorus offers a unique platform with equidistant sequence of spin-up and spin-down states for exploring the quantum spintronic.

\end{abstract}


\maketitle


Few-layer black phosphorus, a two-dimensional semiconductor with a tunable direct band gap, is of a growing importance not only for the potential technological applications but also for the fundamental condensed matter studies\cite{xia2014rediscovering, li2014black, castellanos2014isolation, qiao2014high, takao1981electronic, ling2015renaissance, yuan2015polarization, buscema2014photovoltaic, wang2015highly, yuan2016quantum, rodin2014strain, buscema2014fast, liu2015semiconducting, low2014tunable, low2014plasmons, churchill2014two}. A layer-controlled small band gap of few-layer black phosphorus being on the order of a few hundreds of $meV$ permits tuning the chemical potential between the conduction and valence bands by employing the field effect and thus enables the ambipolar operation of few-layer black phosphorus-based devices\cite{takao1981electronic, akahama1983electrical, tran2014layer}. The development of device fabrication techniques accompanied with the improvement of black phosphorus (BP) crystal quality has experienced a tremendous progress\cite{gillgren2014gate, chen2015high, li2015quantum, long2016type, tayari2015two}. The hole mobility increased from a few hundred to a few thousands $cm^2V^{-1}s^{-1}$\cite{li2016quantum}. A further boost of the hole mobility up to 45,000 $cm^2V^{-1}s^{-1}$ has been made when a BP flake is encapsulated between hexagonal boron nitride (h-BN) in vacuum\cite{long2016achieving}. This high carrier mobility enabled the observation of Shubnikov-de Haas oscillations and the quantum Hall effect in a hole-doped ($p$-type) BP. In spite of these achievements, the performance of electron-doped ($n$-type) BP-based devices remains poor and therefore the electron quantum transport is less explored. According to the theoretical models, the electrons in BP ought to exhibit distinguishable many-body phenomena due to the expected large electron mass\cite{qiao2014high}. Hence, there is a quest to improve the quality of electron-doped black phosphorus crystal and the performance of the devices made thereof, which will eventually lead to the emergence of new quantum phenomena in $n$-type black phosphorus. Here, we take advantage of both encapsulating the BP flakes with h-BN in vacuum conditions and the field effect to control the charge carrier type. The ambipolar device operation demonstrates unambiguously the quantum Hall effect for holes and exhibits clear signatures of quantum Hall effect for electrons.

\section{Transport characteristics of ambipolar BP devices}
\begin{figure}
\includegraphics[scale=1]{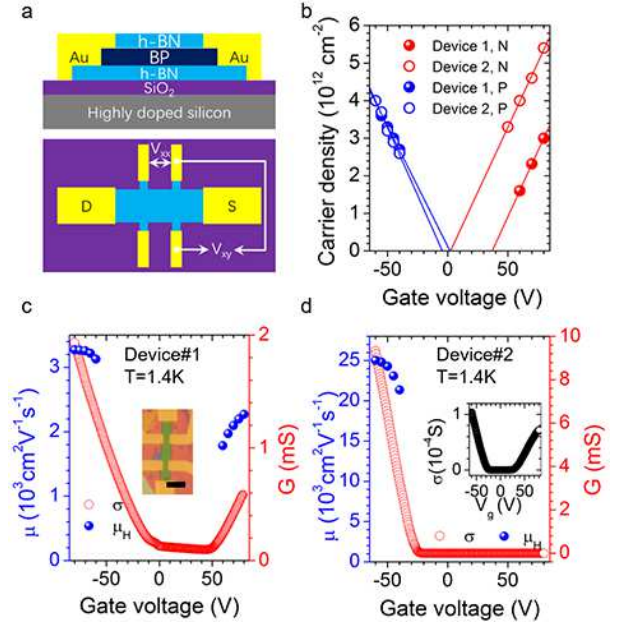}
\caption{\textbf{Transport characteristics of ambipolar black phosphorus field-effect devices.}  (a) Schematic view of BP FETs and measurement configuration. (b) The hole (red) and electron (blue) concentrations obtained from the oscillation periods at varying gate voltages. (c) and (d) Conductance-gate voltage characteristics (red) of device$\sharp$1 (c) and device$\sharp$2 (d) at cryogenic temperature (1.4 K) measured through four terminal configuration. The blue solid dots represent the Hall mobility. The inset of (c) is the micro optical photo of device$\sharp$1. The scale bar is 5 um. The inset of (d) shows the conductance of device$\sharp$2 measured from two terminal configuration.
\label{transport}}
 
\end{figure}

Figure \ref{transport}a displays the schematic side and top views of the hallbar devices fabricated from h-BN/BP/h-BN heterostructure. The h-BN/BP/h-BN heterostructure is assembled in vacuum to reduce the surface absorption and to minimize the amount of impurities on the BP interfaces\cite{long2016achieving}. The Ohmic contacts are made of either chromium or titanium because of their work functions being close to conduction and valence band edges of black phosphorus, respectively\cite{long2016type, perello2015high, cai2014layer}. All the conductance channels are along X direction (Supplementary materials). The electrical measurement are performed with standard lock-in technique(Excitation frequency: 4.579Hz). Device$\sharp$1 has Ti as the Ohmic contact metal, whereas device$\sharp$2 has Cr as the Ohmic contact metal (Supplementary materials). The devices show distinct transport characteristics with a controlled charge carrier type, and their high performance are reproducibly demonstrated on several devices\cite{long2016type}. Figure \ref{transport}b shows the dependence of the charge carrier density on the gate voltage applied between the conducting substrate acting as a gate electrode and one of the Ohmic contacts. More importantly, it manifests a good tuning capability between the electron and hole charge carrier types. Figure \ref{transport}c depicts the four-terminal conductance $G$ of device$\sharp$1 as a function of the gate voltage $V_g$ at temperature $T=1.4K$. The conductance increase for positive and negative gate voltages is accompanied with the increasing Hall mobility $\mu_H=\frac{G}{p(\textrm{or} \: n)\cdot e} \frac{L}{W}$. Here $L$ and $W$ are length and width of the device, respectively, and hole (electron) density $p(n)$ is estimated from the Hall effect for selected gate voltage values.  The Hall mobility for holes and electrons reach 3150 $cm^2V^{-1}s^{-1}$ and 2150 $cm^2V^{-1}s^{-1}$ at $T=1.4K$, respectively. Figure \ref{transport}d summarizes the field effect characteristics of the device$\sharp$2 fabricated with Cr contacts. The transport characteristics measured in a four-terminal configuration shows a typical $p$-type unipolar conductance. The channel conductance increases with the decreasing gate voltage and the Hall mobility reaches  25600 $cm^2V^{-1}s^{-1}$ at $T=1.4K$ - comparable with the previous report\cite{long2016achieving}. For a positive gate voltage, the measurement of $V_{xx}$ fails likely due to the failure of an Ohmic contact. Note, that the energy difference between the work function of Cr and the Fermi level of black phosphorus is rather large for the $n$-type channel. Nonetheless, the device can be characterized in a two-point measurement, the inset of Figure \ref{transport}d depicts the two-point conductance.  The field effect mobility reaches 850 $cm^2V^{-1}s^{-1}$ and 420 $cm^2V^{-1}s^{-1}$ for holes and electrons, respectively. These values are obviously underestimated due to the contact resistance. For holes, the Hall mobility obtained in two-point and four-point measurements differs by factor of 50. Thus, assuming the same factor of underestimation for electrons, we can estimate the electron Hall mobility to be on the order 20,000 $cm^2V^{-1}s^{-1}$. A low Hall mobility in device$\sharp$1 is likely caused by the enhanced scattering rate from the impurities and imperfections of the Si substrate. Indeed, the bottom h-BN layer of device$\sharp$1 is only 4.6 $nm$ and is smaller than 12.3 $nm$ thick h-BN layer in device$\sharp$2. Thus, the charge carrier transport in device$\sharp$1 is stronger affected by the scattering centers on the Si substrate than device$\sharp$2.

\begin{figure*}
\includegraphics[scale=1]{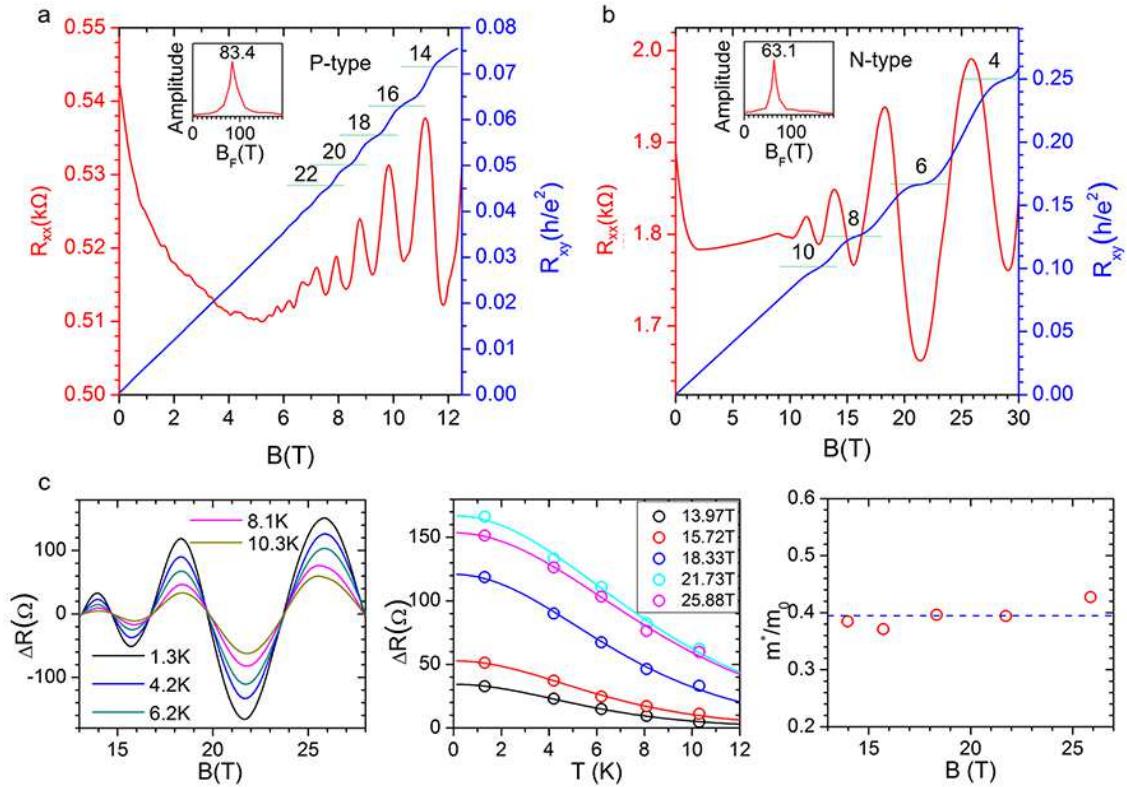}
\caption{\textbf{Quantum Hall effect in ambipolar black phosphorus.} (a) (b) Hall resistance $R_{xy}$ and magnetoresistance $R_{xx}$ as functions of magnetic field B at temperature $T=1.4K$ and the gate voltage of $V_g=-60V$ and $80V$, respectively. The integer number and short dashed green lines marked the filling factors $v$. (c) Left panel: The oscillation components of $n$-type black phosphorus are damped with the increasing temperature. Middle panel: The temperature damping of oscillation components are fitted by the Lifshitz-Kosevitch formula. The solid lines represent the fitting results. Right panel: Effective mass of $n$-type black phosphorus obtained from the fitting results shown in the middle panel.
\label{qheamb}}
\end{figure*}

\section{Ambipolar magnetotransport in BP devices}
We now turn our attention to demonstrate the ambipolar quantum transport characteristics of both devices in the magnetic field. Figure \ref{qheamb}a displays the dependence of $R_{xx}$ and $R_{xy}$ on the magnetic field for hole-type carriers that is obtained upon the application of a backgate voltage $-60V$ on device$\sharp$1. The magnetotransport for electron-type carriers is presented in Figure \ref{qheamb}b when a backgate voltage of $+80V$ is applied. The onset of $R_{xx}$ oscillation at $5T$ for holes and $10T$ for electrons correspond to a quantum mobility of 2000 $cm^2V^{-1}s^{-1}$ and 1000 $cm^2V^{-1}s^{-1}$, respectively. The values approximate the Hall mobility value (Figure \ref{transport}c) and indicate that the large angle scattering event in this device dominate, likely due to a thin bottom h-BN flakes over the small angle scattering  process in this device\cite{long2016achieving, coleridge1991small, piot2005quantum, hwang2008single,knap2004spin}. The insets in Figure \ref{qheamb}a and Figure \ref{qheamb}b show the fast Fourier transformation (FFT) of $R_{xx}$ as a function of $1/B$. The peaks of FFT confirms a periodic oscillation of $R_{xx}$ caused by the formation of Landau levels in the high magnetic field. The carrier concentrations extracted from the FFT frequencies agree well with those determined by the Hall resistance manifesting the absence of another parallel conducting channel in both $n$ and $p$-type BP. At high magnetic field, $R_{xy}$ exhibits the plateau formations corresponding to the quantum Hall resistance $h/ve^2$, where $v$ is an integer number. These plateaus align well with minimas of $R_{xx}$. Thus the magnetotransport signals the quantum Hall effect both for $n$-type and $p$-type conductance in BP device. The left panel of Figure \ref{qheamb}c shows the temperature dependence of $R_{xx}$ oscillatory component for the $n$-type conduction and reveals the damping of the oscillation amplitude with temperature. This damping behavior follows the Lifshitz-Kosevitch formalism $\Delta R \sim \frac{\lambda(T)}{sinh(\lambda(T))} $ , where $\lambda (T) = 2 \pi^2 k_B T m^{\ast} / \hbar e B$ is the temperature damping factor, and allows the estimation of the electron effective mass\cite{shoenberg2009magnetic}. The middle panel of Figure \ref{qheamb}c shows a fairly good fitting of oscillation amplitude damping using the Lifshitz-Kosevitch formalism. The extracted electron effective mass is $0.39\pm0.03 m_0$ and is independent of the magnetic field as shown in Figure \ref{qheamb}c right. This value is in consistence with previous reports\cite{li2015quantum} and the theoretical predications\cite{qiao2014high}.

\paragraph{} The magnetotransport for electron- and hole-type charge carriers shows the conductance quantization at only even filling factors, which suggests that the spin degree of freedom is not resolved even if the magnetic field is high. The discussion of the transport characteristics of device$\sharp$2 presented below may suggest that the high disorder of device$\sharp$1 – reflected in a rather low Hall mobility and a high magnetic field for $R_{xx}$ oscillation onset- can likely account for not observing the lifting of the spin degeneracy. Therefore, this device does not allow to access the spin properties of electrons in black phosphorus.

\begin{figure*}
\includegraphics[scale=1]{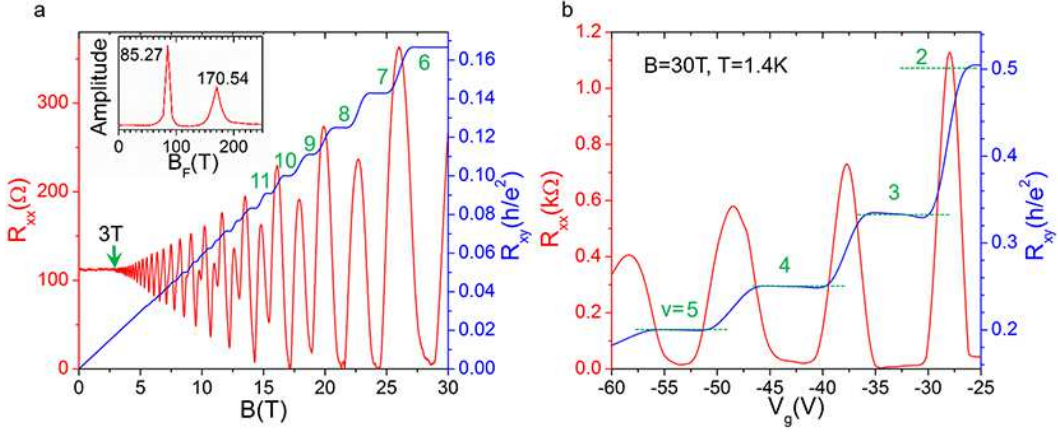}
\caption{\textbf{Quantum Hall effect in $p$-type few layer black phosphorus.} (a) Hall resistance $R_{xy}$ (red) and magnetoresistance $R_{xx}$ (blue) depend on perpendicular magnetic field B at gate voltage $V_g=-60 V$ and temperature $T=1.4 K$. The onset magnetic field is 3T. The inset displays the FFT results of    $R_{xx}$ verses $1/B$. (b) $R_{xy}$ and $R_{xx}$ as functions of gate voltages at $T=1.4K$ under a magnetic field of  $B=30T$. The integer number in (a) and (b) indicate the filling factors $v$ at quantized $R_{xy}=h/ve^2$.
\label{qhep}}
\end{figure*}

\begin{figure*}
\includegraphics[scale=1]{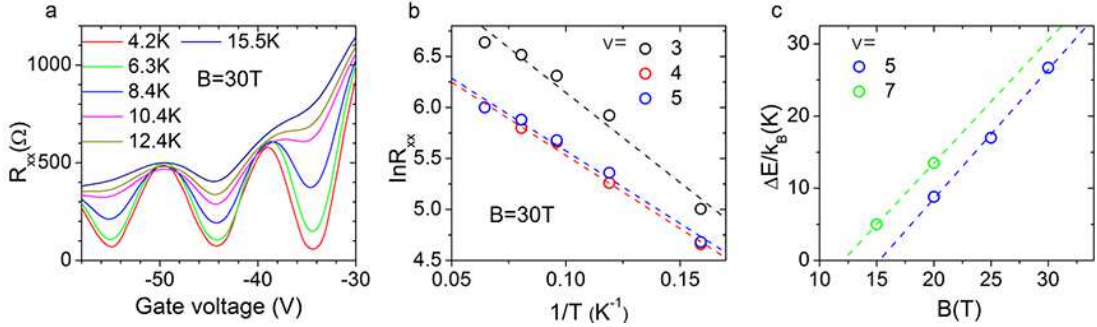}
\caption{\textbf{Excitation energy and Lande g-factor.} (a) $R_{xx}$ depends on gate voltages under $B=30T$ at varying temperatures. (b) ln$R_{xx}$ versus 1/T under different magnetic fields. The solid lines represent the fitting results of Boltzman's law.(c) Excitation energies corresponding to different filling factors obtained from the fitting results shown in (b) as functions of magnetic.
\label{excitation}}
\end{figure*}

\paragraph{} Device$\sharp$2 fabricated with chromium Ohmic contacts has a much higher mobility than observed in device$\sharp$1 suggesting that the level of the disorder is lower and now we turn our attentional to device$\sharp$2. The magnetotransport characteristics for holes in the magnetic field up to $30T$ are shown in Figure \ref{qhep}a. Here the back gate voltage is $-60V$ and the temperature is $T=1.4K$. At low magnetic field ($B<3T$), $R_{xx}$ exhibits nearly zero magnetoresistance indicating the bare existence of localizations in the sample and demonstrating the unprecedented quality of the BP device. The onset of $R_{xx}$ oscillation starts at $3T$ with the sequence of only even Landau level filling factors. This corresponds to a quantum mobility  of 3300 $cm^2V^{-1}s^{-1}$ which is about 8 times lower than Hall mobility. This suggests that the small angle scattering over remote charge impurities dominates over the large angle scattering due to a thicker h-BN bottom layer in device in device$\sharp$2 than in device$\sharp$1.  The splitting of $R_{xx}$ oscillations occurs at $10T$ and develops rapidly in the magnetic field demonstrating the impact of exchange interactions on the splitting which is associated with lifting the spin degeneracy. The Fourier transformation of periodic $R_{xx}$ oscillations on $1/B$-axis show two spectra line at $85T$ and $170T$ (inset of Figure \ref{qhep}a) and thus confirms the lifting of spin degeneracy. When magnetic field reaches 15T, the zero resistance of $R_{xx}$ as well as the quantized $R_{xy}$ plateaus clearly establish the realization of the quantum Hall effect in $p$-type BP device\cite{stone1992quantum}. By sweeping the gate voltage at fixed magnetic field $B=30T$ one gains an access to lower filling factors. Figure \ref{qhep}b reveals a set of $R_{xy}$ plateaus at values $h/e^2v$, where $v$ assumes an integer number from 2 to 5,  accompanied with the zero resistance $R_{xx}$ states. Thus, the Landau level filling factor $v$ can be unequivocally assigned to each plateau. The quantum Hall effect observation at both even and odd filling factors confirms the lifting of spin degeneracy at high magnetic field.

\paragraph{} The zero-resistance $R_{xx}$ states demonstrate well-developed quantum Hall states hence enabled the quantitatively study of the thermal activation energy in our BP device.  Figure \ref{excitation}a shows the thermal activation of $R_{xx}$ at $B=30T$. Figure \ref{excitation}b displays $R_{xx}^{min}$ at three integer Landau level filling factors as a function of $1/T$. The solid lines represent the fitting results according to the  Boltzmann law $R_{xx}^{min} \sim exp(-\Delta E /2k_BT)$, where $\Delta E$ is the activation energy. For black phosphorus the odd filling factors are associated with the Zeeman gap and therefore the activation energy can be expressed as $\Delta E_{odd} = g_Z \mu_B B - \Gamma $, where $\Gamma$ is the Landau level broadening. The activation energy at filling factor $v=5$ and $v=7$ depend linearly on the magnetic field $B$. Thus, since both filling factors are associated with the Zeeman gap, the Lande g-factor is estimated to be $2.5\pm0.3$ and $2.7\pm0.3$  for $v=7$ and $v=5$ in hole doped BP, respectively. These values are in good agreement with our previously reported g-factor value obtained through the Landau level coincidence technique\cite{long2016achieving}.

\begin{figure}
\includegraphics[scale=1]{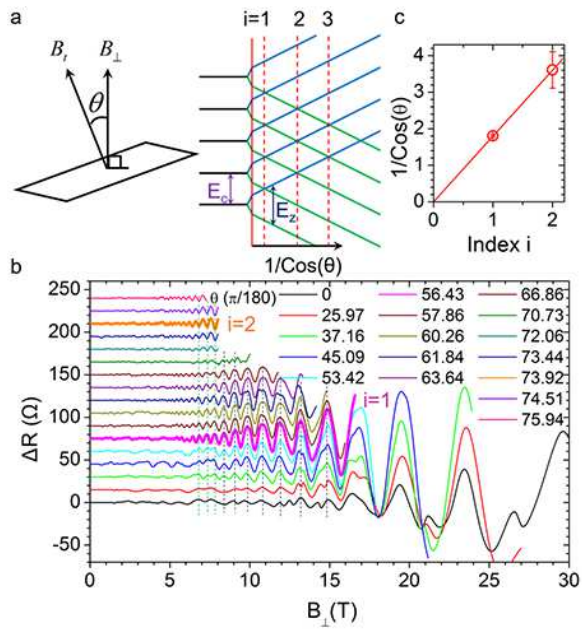}
\caption{\textbf{Landau level crossing of electrons under tilted magnetic field.} 
(a) Left Panel: The configuration of tilt angle $\theta$ between total magnetic field $B_T$ and perpendicular magnetic $B_\perp$. Right Panel: Schematic fan diagram showing spin degeneracy lifted Landau levels evolving with tilt angles. The purple arrow indicates cyclotron energy $E_c=\hbar\omega_c$ and the dark blue arrow represent the Zeeman energy $E_z=g\mu_B B$. The states represented by green solid lines are occupied by spin-up electrons while those represented by blue solid lines are occupied by spin down electrons. The vertical solid red line shows the case of $\theta=0$ and the three vertical dashed red lines indicate the first, second and third coincidence angles, respectively.  (b) Magnetoresistance plotted as a function of  for different tilt angles with fixed gate voltage $V_g=80V$ at cryogenic temperature $T=1.4K$. Coincidences at $i=1$ and $2$ are emphasized as bold lines. (c) $1/Cos(\theta)$ of the identified coincidence angles as a function of i. The solid line represents the linear fitting result indicating a spin-suspecbility $\chi_S$ of 1.1$\pm$ 0.03.
\label{LLC}}
\end{figure}

\paragraph{} Now we draw our attention to the quantum transport characteristics on the electron side. Although the transport characteristics cannot be obtained in the four-point measurements, the two-point magnetotransport characteristics display the oscillations.
The bottom line of figure \ref{LLC}b exemplifies the oscillation component of two-point magnetoresistance $\Delta R$ with $V_g=80V$ and $T=1.4K$ at zero tilt angle. The charge carrier density is determined from the period of the quantum oscillations. 
The minima of oscillations correspond to the integer filling factors and the splitting of quantum oscillations starts at $12T$. This splitting signals the lifting of spin degeneracy, which hence opens the access to explore the spin properties of electrons in black phosphorus. We now utilize this fact to measure the electron spin susceptibility by bringing the spin-resolved Landau level to the coincidence – so-called coincidence technique\cite{schumacher1998anomalous, kozuka2012single, maryenko2014polarization, tsukazaki2008spin, xu2017odd, long2016achieving, fang1968effects, Nicholas1988exchange}. This method profits from the fact that the Zeeman energy $E_Z = g \mu_B B_T$ is given by the total magnetic field $B_T$, while the cyclotron energy $E_c= \frac{\hbar e B_\perp}{m^{\ast}}$ scales with the magnetic field component normal to the the sample plane. Thus $E_c$ and $E_Z$ can be controlled individually by tilting the sample in magnetic field. At certain angle $\theta_i$, the Zeeman energy is an integer of cyclotron energy, i.e. $E_Z=iE_c$, the spin resolved Landau levels overlap as shown in Figure \ref{LLC}a. The spin resolved Landau levels cross each other and $\theta_i$  is the $i^{th}$ coincidence angle. The spin-susceptibility $\chi_s$ can be extracted from the coincidence angle condition $\chi_s =g^{\ast} m^{\ast} =2icos(\theta_i)$. Figure \ref{LLC}b displays the evolution of the oscillatory part of the two-point resistance under a constant $B_\perp$ as $B_T$ and/or tilt angle increases. A coincidence event in the experiment is marked by the vanishing of quantum oscillation components of quantum states at even/odd filling factors i.e. the disappearance of the splitting features. As shown in figure \ref{LLC}b, when the tilt angle approaches $\theta_1=56.4^{\circ}$  even-$v$ $\Delta R$ deeps gradually become weak and finally disappear at $\theta_1$. When the tilt angle increases beyond  $\theta_1$ the deep feature reappears. The same evolution takes place at $\theta_2=73.9^{\circ}$, but  with a large uncertainty.The oscillation amplitudes reaches its maxima at both $\theta_1$ and $\theta_2$ further confirming the coincidence events at these two angles\cite{kurganova2011spin}. The coincidence events at $i=1$ and $i=2$ are emphasized by representing $\Delta R$ with bold lines in figure \ref{LLC}b. Figure \ref{LLC}c plots $1/cos(\theta_i)$ versus $i$. The linear fit of this dependence yields the spin susceptibility for $n$-type BP $\chi_s =1.10 \pm 0.03$. This value of $\chi_s$ illustrates an energetically equidistant arrangement of spin-resolved Landau levels for high filling factors ($v>15$) at zero tilt angle which makes $n$-type BP an ideal platform to explore spin-related phenomena like quantum-spintronic. Considering the electron effective mass $m^{\ast} =0.39 m_0$, the Lande g-factor for $n$-type conducting BP is determined to be $g^{\ast} = 2.8 \pm 0.2$. The enhanced Lande g-factors have been observed in 2DEGs/2DHGs and can be ascribed to the exchange interactions\cite{Nicholas1988exchange, kurganova2011spin, shashkin2001indication, zhu2003spin, vakili2004spin, gokmen2007spin}. This value in $n$-type BP is slightly larger than that in $p$-type conducting BP\cite{long2016achieving}, which might be ascribed to the enhanced electron-electron interactions due to large effective mass of electron\cite{zhou2017effective, girvin2000spin,falson2015even}. The interaction strength is gauged by the ratio of the Coulomb energy to the kinetic energy $r_s=\frac{m^\ast}{\sqrt{\pi n (\textrm{or}\: p)} a_B \kappa m_0}$, where n (or p) is carrier density, $a_B$ is the Bohr radius, and $\kappa$ is the dielectric constant of surrounding medium\cite{movva2017density, maryenko2015spin}.

\paragraph{} In summary, this study demonstrates the quantum Hall effect in the ambipolar operation of the  black phosphorus field effect devices. For hole-doped black phosphorus, we demonstrate the unprecedented quality of the device showing no localization effects around zero field. Furthermore, we measured the Lande g-factor by analyzing the excitation gap at odd filling factors. Its value is consistent with the previous report. For black phosphorus with electron-type carriers, The quantum Hall effect is observed and the Landau level crossing has been studied using the standard coincidence technique. The determined electron mass and Lande g-factor illustrate a equidistant sequence of spin-up and spin-down states makes n-type BP an ideal model system for quantum-spintronic applications.

\paragraph{}  \textbf{Acknowledgements} 
\newline We acknowledge the support of the HFML-RU/FOM, member of the European Magnetic Field Laboratory (EMFL).
\newline Financial support from the Research Grants Council of Hong Kong (Project Nos. 16302215, HKU9/CRF/13G, 604112, and N\_HKUST613/12), Guangdong-Hong Kong joint innovation project (Grant 2016A050503012) and technical support of the Raith-HKUST Nanotechnology Laboratory for the electron-beam lithography facility at MCPF are hereby acknowledged.

\paragraph{} \textbf{Competing financial interests}
\newline The authors declare no competing financial interests.

\paragraph{} \textbf{Author contributions}
\newline G. L. and N. W. conceived the project. G. L. fabricated the devices and performed electronic transport measurements with the help of S. P. and U. Z. G. L., D. M., S. p., U. Z., and N. W. analyzed the data and wrote the manuscript. Other authors provided technical assistance in the project.

\bibliography{QHE_BP}

\begin{thebibliography}{50}%
\makeatletter
\providecommand \@ifxundefined [1]{%
 \@ifx{#1\undefined}
}%
\providecommand \@ifnum [1]{%
 \ifnum #1\expandafter \@firstoftwo
 \else \expandafter \@secondoftwo
 \fi
}%
\providecommand \@ifx [1]{%
 \ifx #1\expandafter \@firstoftwo
 \else \expandafter \@secondoftwo
 \fi
}%
\providecommand \natexlab [1]{#1}%
\providecommand \enquote  [1]{``#1''}%
\providecommand \bibnamefont  [1]{#1}%
\providecommand \bibfnamefont [1]{#1}%
\providecommand \citenamefont [1]{#1}%
\providecommand \href@noop [0]{\@secondoftwo}%
\providecommand \href [0]{\begingroup \@sanitize@url \@href}%
\providecommand \@href[1]{\@@startlink{#1}\@@href}%
\providecommand \@@href[1]{\endgroup#1\@@endlink}%
\providecommand \@sanitize@url [0]{\catcode `\\12\catcode `\$12\catcode
  `\&12\catcode `\#12\catcode `\^12\catcode `\_12\catcode `\%12\relax}%
\providecommand \@@startlink[1]{}%
\providecommand \@@endlink[0]{}%
\providecommand \url  [0]{\begingroup\@sanitize@url \@url }%
\providecommand \@url [1]{\endgroup\@href {#1}{\urlprefix }}%
\providecommand \urlprefix  [0]{URL }%
\providecommand \Eprint [0]{\href }%
\providecommand \doibase [0]{http://dx.doi.org/}%
\providecommand \selectlanguage [0]{\@gobble}%
\providecommand \bibinfo  [0]{\@secondoftwo}%
\providecommand \bibfield  [0]{\@secondoftwo}%
\providecommand \translation [1]{[#1]}%
\providecommand \BibitemOpen [0]{}%
\providecommand \bibitemStop [0]{}%
\providecommand \bibitemNoStop [0]{.\EOS\space}%
\providecommand \EOS [0]{\spacefactor3000\relax}%
\providecommand \BibitemShut  [1]{\csname bibitem#1\endcsname}%
\let\auto@bib@innerbib\@empty
\bibitem [{\citenamefont {Xia}\ \emph {et~al.}(2014)\citenamefont {Xia},
  \citenamefont {Wang},\ and\ \citenamefont {Jia}}]{xia2014rediscovering}%
  \BibitemOpen
  \bibfield  {author} {\bibinfo {author} {\bibfnamefont {F.}~\bibnamefont
  {Xia}}, \bibinfo {author} {\bibfnamefont {H.}~\bibnamefont {Wang}}, \ and\
  \bibinfo {author} {\bibfnamefont {Y.}~\bibnamefont {Jia}},\ }\href@noop {}
  {\bibfield  {journal} {\bibinfo  {journal} {Nature communications}\ }\textbf
  {\bibinfo {volume} {5}},\ \bibinfo {pages} {4458} (\bibinfo {year}
  {2014})}\BibitemShut {NoStop}%
\bibitem [{\citenamefont {Li}\ \emph {et~al.}(2014)\citenamefont {Li},
  \citenamefont {Yu}, \citenamefont {Ye}, \citenamefont {Ge}, \citenamefont
  {Ou}, \citenamefont {Wu}, \citenamefont {Feng}, \citenamefont {Chen},\ and\
  \citenamefont {Zhang}}]{li2014black}%
  \BibitemOpen
  \bibfield  {author} {\bibinfo {author} {\bibfnamefont {L.}~\bibnamefont
  {Li}}, \bibinfo {author} {\bibfnamefont {Y.}~\bibnamefont {Yu}}, \bibinfo
  {author} {\bibfnamefont {G.~J.}\ \bibnamefont {Ye}}, \bibinfo {author}
  {\bibfnamefont {Q.}~\bibnamefont {Ge}}, \bibinfo {author} {\bibfnamefont
  {X.}~\bibnamefont {Ou}}, \bibinfo {author} {\bibfnamefont {H.}~\bibnamefont
  {Wu}}, \bibinfo {author} {\bibfnamefont {D.}~\bibnamefont {Feng}}, \bibinfo
  {author} {\bibfnamefont {X.~H.}\ \bibnamefont {Chen}}, \ and\ \bibinfo
  {author} {\bibfnamefont {Y.}~\bibnamefont {Zhang}},\ }\href@noop {}
  {\bibfield  {journal} {\bibinfo  {journal} {Nature nanotechnology}\ }\textbf
  {\bibinfo {volume} {9}},\ \bibinfo {pages} {372} (\bibinfo {year}
  {2014})}\BibitemShut {NoStop}%
\bibitem [{\citenamefont {Castellanos-Gomez}\ \emph {et~al.}(2014)\citenamefont
  {Castellanos-Gomez}, \citenamefont {Vicarelli}, \citenamefont {Prada},
  \citenamefont {Island}, \citenamefont {Narasimha-Acharya}, \citenamefont
  {Blanter}, \citenamefont {Groenendijk}, \citenamefont {Buscema},
  \citenamefont {Steele}, \citenamefont {Alvarez} \emph
  {et~al.}}]{castellanos2014isolation}%
  \BibitemOpen
  \bibfield  {author} {\bibinfo {author} {\bibfnamefont {A.}~\bibnamefont
  {Castellanos-Gomez}}, \bibinfo {author} {\bibfnamefont {L.}~\bibnamefont
  {Vicarelli}}, \bibinfo {author} {\bibfnamefont {E.}~\bibnamefont {Prada}},
  \bibinfo {author} {\bibfnamefont {J.~O.}\ \bibnamefont {Island}}, \bibinfo
  {author} {\bibfnamefont {K.}~\bibnamefont {Narasimha-Acharya}}, \bibinfo
  {author} {\bibfnamefont {S.~I.}\ \bibnamefont {Blanter}}, \bibinfo {author}
  {\bibfnamefont {D.~J.}\ \bibnamefont {Groenendijk}}, \bibinfo {author}
  {\bibfnamefont {M.}~\bibnamefont {Buscema}}, \bibinfo {author} {\bibfnamefont
  {G.~A.}\ \bibnamefont {Steele}}, \bibinfo {author} {\bibfnamefont
  {J.}~\bibnamefont {Alvarez}},  \emph {et~al.},\ }\href@noop {} {\bibfield
  {journal} {\bibinfo  {journal} {2D Materials}\ }\textbf {\bibinfo {volume}
  {1}},\ \bibinfo {pages} {025001} (\bibinfo {year} {2014})}\BibitemShut
  {NoStop}%
\bibitem [{\citenamefont {Qiao}\ \emph {et~al.}(2014)\citenamefont {Qiao},
  \citenamefont {Kong}, \citenamefont {Hu}, \citenamefont {Yang},\ and\
  \citenamefont {Ji}}]{qiao2014high}%
  \BibitemOpen
  \bibfield  {author} {\bibinfo {author} {\bibfnamefont {J.}~\bibnamefont
  {Qiao}}, \bibinfo {author} {\bibfnamefont {X.}~\bibnamefont {Kong}}, \bibinfo
  {author} {\bibfnamefont {Z.-X.}\ \bibnamefont {Hu}}, \bibinfo {author}
  {\bibfnamefont {F.}~\bibnamefont {Yang}}, \ and\ \bibinfo {author}
  {\bibfnamefont {W.}~\bibnamefont {Ji}},\ }\href@noop {} {\bibfield  {journal}
  {\bibinfo  {journal} {Nature communications}\ }\textbf {\bibinfo {volume}
  {5}},\ \bibinfo {pages} {4475} (\bibinfo {year} {2014})}\BibitemShut
  {NoStop}%
\bibitem [{\citenamefont {Takao}\ \emph {et~al.}(1981)\citenamefont {Takao},
  \citenamefont {Asahina},\ and\ \citenamefont {Morita}}]{takao1981electronic}%
  \BibitemOpen
  \bibfield  {author} {\bibinfo {author} {\bibfnamefont {Y.}~\bibnamefont
  {Takao}}, \bibinfo {author} {\bibfnamefont {H.}~\bibnamefont {Asahina}}, \
  and\ \bibinfo {author} {\bibfnamefont {A.}~\bibnamefont {Morita}},\
  }\href@noop {} {\bibfield  {journal} {\bibinfo  {journal} {Journal of the
  Physical Society of Japan}\ }\textbf {\bibinfo {volume} {50}},\ \bibinfo
  {pages} {3362} (\bibinfo {year} {1981})}\BibitemShut {NoStop}%
\bibitem [{\citenamefont {Ling}\ \emph {et~al.}(2015)\citenamefont {Ling},
  \citenamefont {Wang}, \citenamefont {Huang}, \citenamefont {Xia},\ and\
  \citenamefont {Dresselhaus}}]{ling2015renaissance}%
  \BibitemOpen
  \bibfield  {author} {\bibinfo {author} {\bibfnamefont {X.}~\bibnamefont
  {Ling}}, \bibinfo {author} {\bibfnamefont {H.}~\bibnamefont {Wang}}, \bibinfo
  {author} {\bibfnamefont {S.}~\bibnamefont {Huang}}, \bibinfo {author}
  {\bibfnamefont {F.}~\bibnamefont {Xia}}, \ and\ \bibinfo {author}
  {\bibfnamefont {M.~S.}\ \bibnamefont {Dresselhaus}},\ }\href@noop {}
  {\bibfield  {journal} {\bibinfo  {journal} {Proceedings of the National
  Academy of Sciences}\ }\textbf {\bibinfo {volume} {112}},\ \bibinfo {pages}
  {4523} (\bibinfo {year} {2015})}\BibitemShut {NoStop}%
\bibitem [{\citenamefont {Yuan}\ \emph {et~al.}(2015)\citenamefont {Yuan},
  \citenamefont {Liu}, \citenamefont {Afshinmanesh}, \citenamefont {Li},
  \citenamefont {Xu}, \citenamefont {Sun}, \citenamefont {Lian}, \citenamefont
  {Curto}, \citenamefont {Ye}, \citenamefont {Hikita} \emph
  {et~al.}}]{yuan2015polarization}%
  \BibitemOpen
  \bibfield  {author} {\bibinfo {author} {\bibfnamefont {H.}~\bibnamefont
  {Yuan}}, \bibinfo {author} {\bibfnamefont {X.}~\bibnamefont {Liu}}, \bibinfo
  {author} {\bibfnamefont {F.}~\bibnamefont {Afshinmanesh}}, \bibinfo {author}
  {\bibfnamefont {W.}~\bibnamefont {Li}}, \bibinfo {author} {\bibfnamefont
  {G.}~\bibnamefont {Xu}}, \bibinfo {author} {\bibfnamefont {J.}~\bibnamefont
  {Sun}}, \bibinfo {author} {\bibfnamefont {B.}~\bibnamefont {Lian}}, \bibinfo
  {author} {\bibfnamefont {A.~G.}\ \bibnamefont {Curto}}, \bibinfo {author}
  {\bibfnamefont {G.}~\bibnamefont {Ye}}, \bibinfo {author} {\bibfnamefont
  {Y.}~\bibnamefont {Hikita}},  \emph {et~al.},\ }\href@noop {} {\bibfield
  {journal} {\bibinfo  {journal} {Nature nanotechnology}\ }\textbf {\bibinfo
  {volume} {10}},\ \bibinfo {pages} {707} (\bibinfo {year} {2015})}\BibitemShut
  {NoStop}%
\bibitem [{\citenamefont {Buscema}\ \emph
  {et~al.}(2014{\natexlab{a}})\citenamefont {Buscema}, \citenamefont
  {Groenendijk}, \citenamefont {Steele}, \citenamefont {Van Der~Zant},\ and\
  \citenamefont {Castellanos-Gomez}}]{buscema2014photovoltaic}%
  \BibitemOpen
  \bibfield  {author} {\bibinfo {author} {\bibfnamefont {M.}~\bibnamefont
  {Buscema}}, \bibinfo {author} {\bibfnamefont {D.~J.}\ \bibnamefont
  {Groenendijk}}, \bibinfo {author} {\bibfnamefont {G.~A.}\ \bibnamefont
  {Steele}}, \bibinfo {author} {\bibfnamefont {H.~S.}\ \bibnamefont {Van
  Der~Zant}}, \ and\ \bibinfo {author} {\bibfnamefont {A.}~\bibnamefont
  {Castellanos-Gomez}},\ }\href@noop {} {\bibfield  {journal} {\bibinfo
  {journal} {Nature Communications}\ }\textbf {\bibinfo {volume} {5}},\
  \bibinfo {pages} {4651} (\bibinfo {year} {2014}{\natexlab{a}})}\BibitemShut
  {NoStop}%
\bibitem [{\citenamefont {Wang}\ \emph {et~al.}(2015)\citenamefont {Wang},
  \citenamefont {Jones}, \citenamefont {Seyler}, \citenamefont {Tran},
  \citenamefont {Jia}, \citenamefont {Zhao}, \citenamefont {Wang},
  \citenamefont {Yang}, \citenamefont {Xu},\ and\ \citenamefont
  {Xia}}]{wang2015highly}%
  \BibitemOpen
  \bibfield  {author} {\bibinfo {author} {\bibfnamefont {X.}~\bibnamefont
  {Wang}}, \bibinfo {author} {\bibfnamefont {A.~M.}\ \bibnamefont {Jones}},
  \bibinfo {author} {\bibfnamefont {K.~L.}\ \bibnamefont {Seyler}}, \bibinfo
  {author} {\bibfnamefont {V.}~\bibnamefont {Tran}}, \bibinfo {author}
  {\bibfnamefont {Y.}~\bibnamefont {Jia}}, \bibinfo {author} {\bibfnamefont
  {H.}~\bibnamefont {Zhao}}, \bibinfo {author} {\bibfnamefont {H.}~\bibnamefont
  {Wang}}, \bibinfo {author} {\bibfnamefont {L.}~\bibnamefont {Yang}}, \bibinfo
  {author} {\bibfnamefont {X.}~\bibnamefont {Xu}}, \ and\ \bibinfo {author}
  {\bibfnamefont {F.}~\bibnamefont {Xia}},\ }\href@noop {} {\bibfield
  {journal} {\bibinfo  {journal} {Nature nanotechnology}\ }\textbf {\bibinfo
  {volume} {10}},\ \bibinfo {pages} {517} (\bibinfo {year} {2015})}\BibitemShut
  {NoStop}%
\bibitem [{\citenamefont {Yuan}\ \emph {et~al.}(2016)\citenamefont {Yuan},
  \citenamefont {van Veen}, \citenamefont {Katsnelson},\ and\ \citenamefont
  {Rold{\'a}n}}]{yuan2016quantum}%
  \BibitemOpen
  \bibfield  {author} {\bibinfo {author} {\bibfnamefont {S.}~\bibnamefont
  {Yuan}}, \bibinfo {author} {\bibfnamefont {E.}~\bibnamefont {van Veen}},
  \bibinfo {author} {\bibfnamefont {M.~I.}\ \bibnamefont {Katsnelson}}, \ and\
  \bibinfo {author} {\bibfnamefont {R.}~\bibnamefont {Rold{\'a}n}},\
  }\href@noop {} {\bibfield  {journal} {\bibinfo  {journal} {Physical Review
  B}\ }\textbf {\bibinfo {volume} {93}},\ \bibinfo {pages} {245433} (\bibinfo
  {year} {2016})}\BibitemShut {NoStop}%
\bibitem [{\citenamefont {Rodin}\ \emph {et~al.}(2014)\citenamefont {Rodin},
  \citenamefont {Carvalho},\ and\ \citenamefont {Neto}}]{rodin2014strain}%
  \BibitemOpen
  \bibfield  {author} {\bibinfo {author} {\bibfnamefont {A.}~\bibnamefont
  {Rodin}}, \bibinfo {author} {\bibfnamefont {A.}~\bibnamefont {Carvalho}}, \
  and\ \bibinfo {author} {\bibfnamefont {A.~C.}\ \bibnamefont {Neto}},\
  }\href@noop {} {\bibfield  {journal} {\bibinfo  {journal} {Physical review
  letters}\ }\textbf {\bibinfo {volume} {112}},\ \bibinfo {pages} {176801}
  (\bibinfo {year} {2014})}\BibitemShut {NoStop}%
\bibitem [{\citenamefont {Buscema}\ \emph
  {et~al.}(2014{\natexlab{b}})\citenamefont {Buscema}, \citenamefont
  {Groenendijk}, \citenamefont {Blanter}, \citenamefont {Steele}, \citenamefont
  {van~der Zant},\ and\ \citenamefont {Castellanos-Gomez}}]{buscema2014fast}%
  \BibitemOpen
  \bibfield  {author} {\bibinfo {author} {\bibfnamefont {M.}~\bibnamefont
  {Buscema}}, \bibinfo {author} {\bibfnamefont {D.~J.}\ \bibnamefont
  {Groenendijk}}, \bibinfo {author} {\bibfnamefont {S.~I.}\ \bibnamefont
  {Blanter}}, \bibinfo {author} {\bibfnamefont {G.~A.}\ \bibnamefont {Steele}},
  \bibinfo {author} {\bibfnamefont {H.~S.}\ \bibnamefont {van~der Zant}}, \
  and\ \bibinfo {author} {\bibfnamefont {A.}~\bibnamefont
  {Castellanos-Gomez}},\ }\href@noop {} {\bibfield  {journal} {\bibinfo
  {journal} {Nano letters}\ }\textbf {\bibinfo {volume} {14}},\ \bibinfo
  {pages} {3347} (\bibinfo {year} {2014}{\natexlab{b}})}\BibitemShut {NoStop}%
\bibitem [{\citenamefont {Liu}\ \emph {et~al.}(2015)\citenamefont {Liu},
  \citenamefont {Du}, \citenamefont {Deng},\ and\ \citenamefont
  {Peide}}]{liu2015semiconducting}%
  \BibitemOpen
  \bibfield  {author} {\bibinfo {author} {\bibfnamefont {H.}~\bibnamefont
  {Liu}}, \bibinfo {author} {\bibfnamefont {Y.}~\bibnamefont {Du}}, \bibinfo
  {author} {\bibfnamefont {Y.}~\bibnamefont {Deng}}, \ and\ \bibinfo {author}
  {\bibfnamefont {D.~Y.}\ \bibnamefont {Peide}},\ }\href@noop {} {\bibfield
  {journal} {\bibinfo  {journal} {Chemical Society Reviews}\ }\textbf {\bibinfo
  {volume} {44}},\ \bibinfo {pages} {2732} (\bibinfo {year}
  {2015})}\BibitemShut {NoStop}%
\bibitem [{\citenamefont {Low}\ \emph {et~al.}(2014{\natexlab{a}})\citenamefont
  {Low}, \citenamefont {Rodin}, \citenamefont {Carvalho}, \citenamefont
  {Jiang}, \citenamefont {Wang}, \citenamefont {Xia},\ and\ \citenamefont
  {Neto}}]{low2014tunable}%
  \BibitemOpen
  \bibfield  {author} {\bibinfo {author} {\bibfnamefont {T.}~\bibnamefont
  {Low}}, \bibinfo {author} {\bibfnamefont {A.}~\bibnamefont {Rodin}}, \bibinfo
  {author} {\bibfnamefont {A.}~\bibnamefont {Carvalho}}, \bibinfo {author}
  {\bibfnamefont {Y.}~\bibnamefont {Jiang}}, \bibinfo {author} {\bibfnamefont
  {H.}~\bibnamefont {Wang}}, \bibinfo {author} {\bibfnamefont {F.}~\bibnamefont
  {Xia}}, \ and\ \bibinfo {author} {\bibfnamefont {A.~C.}\ \bibnamefont
  {Neto}},\ }\href@noop {} {\bibfield  {journal} {\bibinfo  {journal} {Physical
  Review B}\ }\textbf {\bibinfo {volume} {90}},\ \bibinfo {pages} {075434}
  (\bibinfo {year} {2014}{\natexlab{a}})}\BibitemShut {NoStop}%
\bibitem [{\citenamefont {Low}\ \emph {et~al.}(2014{\natexlab{b}})\citenamefont
  {Low}, \citenamefont {Rold{\'a}n}, \citenamefont {Wang}, \citenamefont {Xia},
  \citenamefont {Avouris}, \citenamefont {Moreno},\ and\ \citenamefont
  {Guinea}}]{low2014plasmons}%
  \BibitemOpen
  \bibfield  {author} {\bibinfo {author} {\bibfnamefont {T.}~\bibnamefont
  {Low}}, \bibinfo {author} {\bibfnamefont {R.}~\bibnamefont {Rold{\'a}n}},
  \bibinfo {author} {\bibfnamefont {H.}~\bibnamefont {Wang}}, \bibinfo {author}
  {\bibfnamefont {F.}~\bibnamefont {Xia}}, \bibinfo {author} {\bibfnamefont
  {P.}~\bibnamefont {Avouris}}, \bibinfo {author} {\bibfnamefont {L.~M.}\
  \bibnamefont {Moreno}}, \ and\ \bibinfo {author} {\bibfnamefont
  {F.}~\bibnamefont {Guinea}},\ }\href@noop {} {\bibfield  {journal} {\bibinfo
  {journal} {Physical review letters}\ }\textbf {\bibinfo {volume} {113}},\
  \bibinfo {pages} {106802} (\bibinfo {year} {2014}{\natexlab{b}})}\BibitemShut
  {NoStop}%
\bibitem [{\citenamefont {Churchill}\ and\ \citenamefont
  {Jarillo-Herrero}(2014)}]{churchill2014two}%
  \BibitemOpen
  \bibfield  {author} {\bibinfo {author} {\bibfnamefont {H.~O.}\ \bibnamefont
  {Churchill}}\ and\ \bibinfo {author} {\bibfnamefont {P.}~\bibnamefont
  {Jarillo-Herrero}},\ }\href@noop {} {\bibfield  {journal} {\bibinfo
  {journal} {Nature nanotechnology}\ }\textbf {\bibinfo {volume} {9}},\
  \bibinfo {pages} {330} (\bibinfo {year} {2014})}\BibitemShut {NoStop}%
\bibitem [{\citenamefont {Akahama}\ \emph {et~al.}(1983)\citenamefont
  {Akahama}, \citenamefont {Endo},\ and\ \citenamefont
  {Narita}}]{akahama1983electrical}%
  \BibitemOpen
  \bibfield  {author} {\bibinfo {author} {\bibfnamefont {Y.}~\bibnamefont
  {Akahama}}, \bibinfo {author} {\bibfnamefont {S.}~\bibnamefont {Endo}}, \
  and\ \bibinfo {author} {\bibfnamefont {S.-i.}\ \bibnamefont {Narita}},\
  }\href@noop {} {\bibfield  {journal} {\bibinfo  {journal} {Journal of the
  Physical Society of Japan}\ }\textbf {\bibinfo {volume} {52}},\ \bibinfo
  {pages} {2148} (\bibinfo {year} {1983})}\BibitemShut {NoStop}%
\bibitem [{\citenamefont {Tran}\ \emph {et~al.}(2014)\citenamefont {Tran},
  \citenamefont {Soklaski}, \citenamefont {Liang},\ and\ \citenamefont
  {Yang}}]{tran2014layer}%
  \BibitemOpen
  \bibfield  {author} {\bibinfo {author} {\bibfnamefont {V.}~\bibnamefont
  {Tran}}, \bibinfo {author} {\bibfnamefont {R.}~\bibnamefont {Soklaski}},
  \bibinfo {author} {\bibfnamefont {Y.}~\bibnamefont {Liang}}, \ and\ \bibinfo
  {author} {\bibfnamefont {L.}~\bibnamefont {Yang}},\ }\href@noop {} {\bibfield
   {journal} {\bibinfo  {journal} {Physical Review B}\ }\textbf {\bibinfo
  {volume} {89}},\ \bibinfo {pages} {235319} (\bibinfo {year}
  {2014})}\BibitemShut {NoStop}%
\bibitem [{\citenamefont {Gillgren}\ \emph {et~al.}(2014)\citenamefont
  {Gillgren}, \citenamefont {Wickramaratne}, \citenamefont {Shi}, \citenamefont
  {Espiritu}, \citenamefont {Yang}, \citenamefont {Hu}, \citenamefont {Wei},
  \citenamefont {Liu}, \citenamefont {Mao}, \citenamefont {Watanabe} \emph
  {et~al.}}]{gillgren2014gate}%
  \BibitemOpen
  \bibfield  {author} {\bibinfo {author} {\bibfnamefont {N.}~\bibnamefont
  {Gillgren}}, \bibinfo {author} {\bibfnamefont {D.}~\bibnamefont
  {Wickramaratne}}, \bibinfo {author} {\bibfnamefont {Y.}~\bibnamefont {Shi}},
  \bibinfo {author} {\bibfnamefont {T.}~\bibnamefont {Espiritu}}, \bibinfo
  {author} {\bibfnamefont {J.}~\bibnamefont {Yang}}, \bibinfo {author}
  {\bibfnamefont {J.}~\bibnamefont {Hu}}, \bibinfo {author} {\bibfnamefont
  {J.}~\bibnamefont {Wei}}, \bibinfo {author} {\bibfnamefont {X.}~\bibnamefont
  {Liu}}, \bibinfo {author} {\bibfnamefont {Z.}~\bibnamefont {Mao}}, \bibinfo
  {author} {\bibfnamefont {K.}~\bibnamefont {Watanabe}},  \emph {et~al.},\
  }\href@noop {} {\bibfield  {journal} {\bibinfo  {journal} {2D Materials}\
  }\textbf {\bibinfo {volume} {2}},\ \bibinfo {pages} {011001} (\bibinfo {year}
  {2014})}\BibitemShut {NoStop}%
\bibitem [{\citenamefont {Chen}\ \emph {et~al.}(2015)\citenamefont {Chen},
  \citenamefont {Wu}, \citenamefont {Wu}, \citenamefont {Han}, \citenamefont
  {Xu}, \citenamefont {Wang}, \citenamefont {Ye}, \citenamefont {Han},
  \citenamefont {He}, \citenamefont {Cai} \emph {et~al.}}]{chen2015high}%
  \BibitemOpen
  \bibfield  {author} {\bibinfo {author} {\bibfnamefont {X.}~\bibnamefont
  {Chen}}, \bibinfo {author} {\bibfnamefont {Y.}~\bibnamefont {Wu}}, \bibinfo
  {author} {\bibfnamefont {Z.}~\bibnamefont {Wu}}, \bibinfo {author}
  {\bibfnamefont {Y.}~\bibnamefont {Han}}, \bibinfo {author} {\bibfnamefont
  {S.}~\bibnamefont {Xu}}, \bibinfo {author} {\bibfnamefont {L.}~\bibnamefont
  {Wang}}, \bibinfo {author} {\bibfnamefont {W.}~\bibnamefont {Ye}}, \bibinfo
  {author} {\bibfnamefont {T.}~\bibnamefont {Han}}, \bibinfo {author}
  {\bibfnamefont {Y.}~\bibnamefont {He}}, \bibinfo {author} {\bibfnamefont
  {Y.}~\bibnamefont {Cai}},  \emph {et~al.},\ }\href@noop {} {\bibfield
  {journal} {\bibinfo  {journal} {Nature communications}\ }\textbf {\bibinfo
  {volume} {6}},\ \bibinfo {pages} {7315} (\bibinfo {year} {2015})}\BibitemShut
  {NoStop}%
\bibitem [{\citenamefont {Li}\ \emph {et~al.}(2015)\citenamefont {Li},
  \citenamefont {Ye}, \citenamefont {Tran}, \citenamefont {Fei}, \citenamefont
  {Chen}, \citenamefont {Wang}, \citenamefont {Wang}, \citenamefont {Watanabe},
  \citenamefont {Taniguchi}, \citenamefont {Yang} \emph
  {et~al.}}]{li2015quantum}%
  \BibitemOpen
  \bibfield  {author} {\bibinfo {author} {\bibfnamefont {L.}~\bibnamefont
  {Li}}, \bibinfo {author} {\bibfnamefont {G.~J.}\ \bibnamefont {Ye}}, \bibinfo
  {author} {\bibfnamefont {V.}~\bibnamefont {Tran}}, \bibinfo {author}
  {\bibfnamefont {R.}~\bibnamefont {Fei}}, \bibinfo {author} {\bibfnamefont
  {G.}~\bibnamefont {Chen}}, \bibinfo {author} {\bibfnamefont {H.}~\bibnamefont
  {Wang}}, \bibinfo {author} {\bibfnamefont {J.}~\bibnamefont {Wang}}, \bibinfo
  {author} {\bibfnamefont {K.}~\bibnamefont {Watanabe}}, \bibinfo {author}
  {\bibfnamefont {T.}~\bibnamefont {Taniguchi}}, \bibinfo {author}
  {\bibfnamefont {L.}~\bibnamefont {Yang}},  \emph {et~al.},\ }\href@noop {}
  {\bibfield  {journal} {\bibinfo  {journal} {Nature nanotechnology}\ }\textbf
  {\bibinfo {volume} {10}},\ \bibinfo {pages} {608} (\bibinfo {year}
  {2015})}\BibitemShut {NoStop}%
\bibitem [{\citenamefont {Long}\ \emph
  {et~al.}(2016{\natexlab{a}})\citenamefont {Long}, \citenamefont {Xu},
  \citenamefont {Shen}, \citenamefont {Hou}, \citenamefont {Wu}, \citenamefont
  {Han}, \citenamefont {Lin}, \citenamefont {Wong}, \citenamefont {Cai},
  \citenamefont {Lortz} \emph {et~al.}}]{long2016type}%
  \BibitemOpen
  \bibfield  {author} {\bibinfo {author} {\bibfnamefont {G.}~\bibnamefont
  {Long}}, \bibinfo {author} {\bibfnamefont {S.}~\bibnamefont {Xu}}, \bibinfo
  {author} {\bibfnamefont {J.}~\bibnamefont {Shen}}, \bibinfo {author}
  {\bibfnamefont {J.}~\bibnamefont {Hou}}, \bibinfo {author} {\bibfnamefont
  {Z.}~\bibnamefont {Wu}}, \bibinfo {author} {\bibfnamefont {T.}~\bibnamefont
  {Han}}, \bibinfo {author} {\bibfnamefont {J.}~\bibnamefont {Lin}}, \bibinfo
  {author} {\bibfnamefont {W.~K.}\ \bibnamefont {Wong}}, \bibinfo {author}
  {\bibfnamefont {Y.}~\bibnamefont {Cai}}, \bibinfo {author} {\bibfnamefont
  {R.}~\bibnamefont {Lortz}},  \emph {et~al.},\ }\href@noop {} {\bibfield
  {journal} {\bibinfo  {journal} {2D Materials}\ }\textbf {\bibinfo {volume}
  {3}},\ \bibinfo {pages} {031001} (\bibinfo {year}
  {2016}{\natexlab{a}})}\BibitemShut {NoStop}%
\bibitem [{\citenamefont {Tayari}\ \emph {et~al.}(2015)\citenamefont {Tayari},
  \citenamefont {Hemsworth}, \citenamefont {Fakih}, \citenamefont {Favron},
  \citenamefont {Gaufres}, \citenamefont {Gervais}, \citenamefont {Martel},\
  and\ \citenamefont {Szkopek}}]{tayari2015two}%
  \BibitemOpen
  \bibfield  {author} {\bibinfo {author} {\bibfnamefont {V.}~\bibnamefont
  {Tayari}}, \bibinfo {author} {\bibfnamefont {N.}~\bibnamefont {Hemsworth}},
  \bibinfo {author} {\bibfnamefont {I.}~\bibnamefont {Fakih}}, \bibinfo
  {author} {\bibfnamefont {A.}~\bibnamefont {Favron}}, \bibinfo {author}
  {\bibfnamefont {E.}~\bibnamefont {Gaufres}}, \bibinfo {author} {\bibfnamefont
  {G.}~\bibnamefont {Gervais}}, \bibinfo {author} {\bibfnamefont
  {R.}~\bibnamefont {Martel}}, \ and\ \bibinfo {author} {\bibfnamefont
  {T.}~\bibnamefont {Szkopek}},\ }\href@noop {} {\bibfield  {journal} {\bibinfo
   {journal} {Nature communications}\ }\textbf {\bibinfo {volume} {6}},\
  \bibinfo {pages} {7702} (\bibinfo {year} {2015})}\BibitemShut {NoStop}%
\bibitem [{\citenamefont {Li}\ \emph {et~al.}(2016)\citenamefont {Li},
  \citenamefont {Yang}, \citenamefont {Ye}, \citenamefont {Zhang},
  \citenamefont {Zhu}, \citenamefont {Lou}, \citenamefont {Zhou}, \citenamefont
  {Li}, \citenamefont {Watanabe}, \citenamefont {Taniguchi} \emph
  {et~al.}}]{li2016quantum}%
  \BibitemOpen
  \bibfield  {author} {\bibinfo {author} {\bibfnamefont {L.}~\bibnamefont
  {Li}}, \bibinfo {author} {\bibfnamefont {F.}~\bibnamefont {Yang}}, \bibinfo
  {author} {\bibfnamefont {G.~J.}\ \bibnamefont {Ye}}, \bibinfo {author}
  {\bibfnamefont {Z.}~\bibnamefont {Zhang}}, \bibinfo {author} {\bibfnamefont
  {Z.}~\bibnamefont {Zhu}}, \bibinfo {author} {\bibfnamefont {W.}~\bibnamefont
  {Lou}}, \bibinfo {author} {\bibfnamefont {X.}~\bibnamefont {Zhou}}, \bibinfo
  {author} {\bibfnamefont {L.}~\bibnamefont {Li}}, \bibinfo {author}
  {\bibfnamefont {K.}~\bibnamefont {Watanabe}}, \bibinfo {author}
  {\bibfnamefont {T.}~\bibnamefont {Taniguchi}},  \emph {et~al.},\ }\href@noop
  {} {\bibfield  {journal} {\bibinfo  {journal} {Nature nanotechnology}\
  }\textbf {\bibinfo {volume} {11}},\ \bibinfo {pages} {593} (\bibinfo {year}
  {2016})}\BibitemShut {NoStop}%
\bibitem [{\citenamefont {Long}\ \emph
  {et~al.}(2016{\natexlab{b}})\citenamefont {Long}, \citenamefont {Maryenko},
  \citenamefont {Shen}, \citenamefont {Xu}, \citenamefont {Hou}, \citenamefont
  {Wu}, \citenamefont {Wong}, \citenamefont {Han}, \citenamefont {Lin},
  \citenamefont {Cai} \emph {et~al.}}]{long2016achieving}%
  \BibitemOpen
  \bibfield  {author} {\bibinfo {author} {\bibfnamefont {G.}~\bibnamefont
  {Long}}, \bibinfo {author} {\bibfnamefont {D.}~\bibnamefont {Maryenko}},
  \bibinfo {author} {\bibfnamefont {J.}~\bibnamefont {Shen}}, \bibinfo {author}
  {\bibfnamefont {S.}~\bibnamefont {Xu}}, \bibinfo {author} {\bibfnamefont
  {J.}~\bibnamefont {Hou}}, \bibinfo {author} {\bibfnamefont {Z.}~\bibnamefont
  {Wu}}, \bibinfo {author} {\bibfnamefont {W.~K.}\ \bibnamefont {Wong}},
  \bibinfo {author} {\bibfnamefont {T.}~\bibnamefont {Han}}, \bibinfo {author}
  {\bibfnamefont {J.}~\bibnamefont {Lin}}, \bibinfo {author} {\bibfnamefont
  {Y.}~\bibnamefont {Cai}},  \emph {et~al.},\ }\href@noop {} {\bibfield
  {journal} {\bibinfo  {journal} {Nano Letters}\ }\textbf {\bibinfo {volume}
  {16 (12)}},\ \bibinfo {pages} {7768} (\bibinfo {year}
  {2016}{\natexlab{b}})}\BibitemShut {NoStop}%
\bibitem [{\citenamefont {Perello}\ \emph {et~al.}(2015)\citenamefont
  {Perello}, \citenamefont {Chae}, \citenamefont {Song},\ and\ \citenamefont
  {Lee}}]{perello2015high}%
  \BibitemOpen
  \bibfield  {author} {\bibinfo {author} {\bibfnamefont {D.~J.}\ \bibnamefont
  {Perello}}, \bibinfo {author} {\bibfnamefont {S.~H.}\ \bibnamefont {Chae}},
  \bibinfo {author} {\bibfnamefont {S.}~\bibnamefont {Song}}, \ and\ \bibinfo
  {author} {\bibfnamefont {Y.~H.}\ \bibnamefont {Lee}},\ }\href@noop {}
  {\bibfield  {journal} {\bibinfo  {journal} {Nature communications}\ }\textbf
  {\bibinfo {volume} {6}},\ \bibinfo {pages} {7809} (\bibinfo {year}
  {2015})}\BibitemShut {NoStop}%
\bibitem [{\citenamefont {Cai}\ \emph {et~al.}(2014)\citenamefont {Cai},
  \citenamefont {Zhang},\ and\ \citenamefont {Zhang}}]{cai2014layer}%
  \BibitemOpen
  \bibfield  {author} {\bibinfo {author} {\bibfnamefont {Y.}~\bibnamefont
  {Cai}}, \bibinfo {author} {\bibfnamefont {G.}~\bibnamefont {Zhang}}, \ and\
  \bibinfo {author} {\bibfnamefont {Y.-W.}\ \bibnamefont {Zhang}},\ }\href@noop
  {} {\bibfield  {journal} {\bibinfo  {journal} {Scientific Reports}\ }\textbf
  {\bibinfo {volume} {4}},\ \bibinfo {pages} {6677} (\bibinfo {year}
  {2014})}\BibitemShut {NoStop}%
\bibitem [{\citenamefont {Coleridge}(1991)}]{coleridge1991small}%
  \BibitemOpen
  \bibfield  {author} {\bibinfo {author} {\bibfnamefont {P.}~\bibnamefont
  {Coleridge}},\ }\href@noop {} {\bibfield  {journal} {\bibinfo  {journal}
  {Physical Review B}\ }\textbf {\bibinfo {volume} {44}},\ \bibinfo {pages}
  {3793} (\bibinfo {year} {1991})}\BibitemShut {NoStop}%
\bibitem [{\citenamefont {Piot}\ \emph {et~al.}(2005)\citenamefont {Piot},
  \citenamefont {Maude}, \citenamefont {Henini}, \citenamefont {Wasilewski},
  \citenamefont {Friedland}, \citenamefont {Hey}, \citenamefont {Ploog},
  \citenamefont {Toropov}, \citenamefont {Airey},\ and\ \citenamefont
  {Hill}}]{piot2005quantum}%
  \BibitemOpen
  \bibfield  {author} {\bibinfo {author} {\bibfnamefont {B.}~\bibnamefont
  {Piot}}, \bibinfo {author} {\bibfnamefont {D.}~\bibnamefont {Maude}},
  \bibinfo {author} {\bibfnamefont {M.}~\bibnamefont {Henini}}, \bibinfo
  {author} {\bibfnamefont {Z.}~\bibnamefont {Wasilewski}}, \bibinfo {author}
  {\bibfnamefont {K.}~\bibnamefont {Friedland}}, \bibinfo {author}
  {\bibfnamefont {R.}~\bibnamefont {Hey}}, \bibinfo {author} {\bibfnamefont
  {K.}~\bibnamefont {Ploog}}, \bibinfo {author} {\bibfnamefont
  {A.}~\bibnamefont {Toropov}}, \bibinfo {author} {\bibfnamefont
  {R.}~\bibnamefont {Airey}}, \ and\ \bibinfo {author} {\bibfnamefont
  {G.}~\bibnamefont {Hill}},\ }\href@noop {} {\bibfield  {journal} {\bibinfo
  {journal} {Physical Review B}\ }\textbf {\bibinfo {volume} {72}},\ \bibinfo
  {pages} {245325} (\bibinfo {year} {2005})}\BibitemShut {NoStop}%
\bibitem [{\citenamefont {Hwang}\ and\ \citenamefont
  {Sarma}(2008)}]{hwang2008single}%
  \BibitemOpen
  \bibfield  {author} {\bibinfo {author} {\bibfnamefont {E.}~\bibnamefont
  {Hwang}}\ and\ \bibinfo {author} {\bibfnamefont {S.~D.}\ \bibnamefont
  {Sarma}},\ }\href@noop {} {\bibfield  {journal} {\bibinfo  {journal}
  {Physical Review B}\ }\textbf {\bibinfo {volume} {77}},\ \bibinfo {pages}
  {195412} (\bibinfo {year} {2008})}\BibitemShut {NoStop}%
\bibitem [{\citenamefont {Knap}\ \emph {et~al.}(2004)\citenamefont {Knap},
  \citenamefont {Fal’ko}, \citenamefont {Frayssinet}, \citenamefont
  {Lorenzini}, \citenamefont {Grandjean}, \citenamefont {Maude}, \citenamefont
  {Karczewski}, \citenamefont {Brandt}, \citenamefont {{\L}usakowski},
  \citenamefont {Grzegory} \emph {et~al.}}]{knap2004spin}%
  \BibitemOpen
  \bibfield  {author} {\bibinfo {author} {\bibfnamefont {W.}~\bibnamefont
  {Knap}}, \bibinfo {author} {\bibfnamefont {V.}~\bibnamefont {Fal’ko}},
  \bibinfo {author} {\bibfnamefont {E.}~\bibnamefont {Frayssinet}}, \bibinfo
  {author} {\bibfnamefont {P.}~\bibnamefont {Lorenzini}}, \bibinfo {author}
  {\bibfnamefont {N.}~\bibnamefont {Grandjean}}, \bibinfo {author}
  {\bibfnamefont {D.}~\bibnamefont {Maude}}, \bibinfo {author} {\bibfnamefont
  {G.}~\bibnamefont {Karczewski}}, \bibinfo {author} {\bibfnamefont
  {B.}~\bibnamefont {Brandt}}, \bibinfo {author} {\bibfnamefont
  {J.}~\bibnamefont {{\L}usakowski}}, \bibinfo {author} {\bibfnamefont
  {I.}~\bibnamefont {Grzegory}},  \emph {et~al.},\ }\href@noop {} {\bibfield
  {journal} {\bibinfo  {journal} {Journal of Physics: Condensed Matter}\
  }\textbf {\bibinfo {volume} {16}},\ \bibinfo {pages} {3421} (\bibinfo {year}
  {2004})}\BibitemShut {NoStop}%
\bibitem [{\citenamefont {Shoenberg}(2009)}]{shoenberg2009magnetic}%
  \BibitemOpen
  \bibfield  {author} {\bibinfo {author} {\bibfnamefont {D.}~\bibnamefont
  {Shoenberg}},\ }\href@noop {} {\emph {\bibinfo {title} {Magnetic oscillations
  in metals}}}\ (\bibinfo  {publisher} {Cambridge University Press},\ \bibinfo
  {year} {2009})\BibitemShut {NoStop}%
\bibitem [{\citenamefont {Stone}(1992)}]{stone1992quantum}%
  \BibitemOpen
  \bibfield  {author} {\bibinfo {author} {\bibfnamefont {M.}~\bibnamefont
  {Stone}},\ }\href@noop {} {\emph {\bibinfo {title} {Quantum Hall Effect}}}\
  (\bibinfo  {publisher} {World Scientific},\ \bibinfo {year}
  {1992})\BibitemShut {NoStop}%
\bibitem [{\citenamefont {Schumacher}\ \emph {et~al.}(1998)\citenamefont
  {Schumacher}, \citenamefont {Nauen}, \citenamefont {Zeitler}, \citenamefont
  {Haug}, \citenamefont {Weitz}, \citenamefont {Jansen},\ and\ \citenamefont
  {Sch{\"a}ffler}}]{schumacher1998anomalous}%
  \BibitemOpen
  \bibfield  {author} {\bibinfo {author} {\bibfnamefont {H.}~\bibnamefont
  {Schumacher}}, \bibinfo {author} {\bibfnamefont {A.}~\bibnamefont {Nauen}},
  \bibinfo {author} {\bibfnamefont {U.}~\bibnamefont {Zeitler}}, \bibinfo
  {author} {\bibfnamefont {R.}~\bibnamefont {Haug}}, \bibinfo {author}
  {\bibfnamefont {P.}~\bibnamefont {Weitz}}, \bibinfo {author} {\bibfnamefont
  {A.}~\bibnamefont {Jansen}}, \ and\ \bibinfo {author} {\bibfnamefont
  {F.}~\bibnamefont {Sch{\"a}ffler}},\ }\href@noop {} {\bibfield  {journal}
  {\bibinfo  {journal} {Physica B: Condensed Matter}\ }\textbf {\bibinfo
  {volume} {256}},\ \bibinfo {pages} {260} (\bibinfo {year}
  {1998})}\BibitemShut {NoStop}%
\bibitem [{\citenamefont {Kozuka}\ \emph {et~al.}(2012)\citenamefont {Kozuka},
  \citenamefont {Tsukazaki}, \citenamefont {Maryenko}, \citenamefont {Falson},
  \citenamefont {Bell}, \citenamefont {Kim}, \citenamefont {Hikita},
  \citenamefont {Hwang},\ and\ \citenamefont {Kawasaki}}]{kozuka2012single}%
  \BibitemOpen
  \bibfield  {author} {\bibinfo {author} {\bibfnamefont {Y.}~\bibnamefont
  {Kozuka}}, \bibinfo {author} {\bibfnamefont {A.}~\bibnamefont {Tsukazaki}},
  \bibinfo {author} {\bibfnamefont {D.}~\bibnamefont {Maryenko}}, \bibinfo
  {author} {\bibfnamefont {J.}~\bibnamefont {Falson}}, \bibinfo {author}
  {\bibfnamefont {C.}~\bibnamefont {Bell}}, \bibinfo {author} {\bibfnamefont
  {M.}~\bibnamefont {Kim}}, \bibinfo {author} {\bibfnamefont {Y.}~\bibnamefont
  {Hikita}}, \bibinfo {author} {\bibfnamefont {H.}~\bibnamefont {Hwang}}, \
  and\ \bibinfo {author} {\bibfnamefont {M.}~\bibnamefont {Kawasaki}},\
  }\href@noop {} {\bibfield  {journal} {\bibinfo  {journal} {Physical Review
  B}\ }\textbf {\bibinfo {volume} {85}},\ \bibinfo {pages} {075302} (\bibinfo
  {year} {2012})}\BibitemShut {NoStop}%
\bibitem [{\citenamefont {Maryenko}\ \emph {et~al.}(2014)\citenamefont
  {Maryenko}, \citenamefont {Falson}, \citenamefont {Kozuka}, \citenamefont
  {Tsukazaki},\ and\ \citenamefont {Kawasaki}}]{maryenko2014polarization}%
  \BibitemOpen
  \bibfield  {author} {\bibinfo {author} {\bibfnamefont {D.}~\bibnamefont
  {Maryenko}}, \bibinfo {author} {\bibfnamefont {J.}~\bibnamefont {Falson}},
  \bibinfo {author} {\bibfnamefont {Y.}~\bibnamefont {Kozuka}}, \bibinfo
  {author} {\bibfnamefont {A.}~\bibnamefont {Tsukazaki}}, \ and\ \bibinfo
  {author} {\bibfnamefont {M.}~\bibnamefont {Kawasaki}},\ }\href@noop {}
  {\bibfield  {journal} {\bibinfo  {journal} {Physical Review B}\ }\textbf
  {\bibinfo {volume} {90}},\ \bibinfo {pages} {245303} (\bibinfo {year}
  {2014})}\BibitemShut {NoStop}%
\bibitem [{\citenamefont {Tsukazaki}\ \emph {et~al.}(2008)\citenamefont
  {Tsukazaki}, \citenamefont {Ohtomo}, \citenamefont {Kawasaki}, \citenamefont
  {Akasaka}, \citenamefont {Yuji}, \citenamefont {Tamura}, \citenamefont
  {Nakahara}, \citenamefont {Tanabe}, \citenamefont {Kamisawa}, \citenamefont
  {Gokmen} \emph {et~al.}}]{tsukazaki2008spin}%
  \BibitemOpen
  \bibfield  {author} {\bibinfo {author} {\bibfnamefont {A.}~\bibnamefont
  {Tsukazaki}}, \bibinfo {author} {\bibfnamefont {A.}~\bibnamefont {Ohtomo}},
  \bibinfo {author} {\bibfnamefont {M.}~\bibnamefont {Kawasaki}}, \bibinfo
  {author} {\bibfnamefont {S.}~\bibnamefont {Akasaka}}, \bibinfo {author}
  {\bibfnamefont {H.}~\bibnamefont {Yuji}}, \bibinfo {author} {\bibfnamefont
  {K.}~\bibnamefont {Tamura}}, \bibinfo {author} {\bibfnamefont
  {K.}~\bibnamefont {Nakahara}}, \bibinfo {author} {\bibfnamefont
  {T.}~\bibnamefont {Tanabe}}, \bibinfo {author} {\bibfnamefont
  {A.}~\bibnamefont {Kamisawa}}, \bibinfo {author} {\bibfnamefont
  {T.}~\bibnamefont {Gokmen}},  \emph {et~al.},\ }\href@noop {} {\bibfield
  {journal} {\bibinfo  {journal} {Physical Review B}\ }\textbf {\bibinfo
  {volume} {78}},\ \bibinfo {pages} {233308} (\bibinfo {year}
  {2008})}\BibitemShut {NoStop}%
\bibitem [{\citenamefont {Xu}\ \emph {et~al.}(2017)\citenamefont {Xu},
  \citenamefont {Shen}, \citenamefont {Long}, \citenamefont {Wu}, \citenamefont
  {Bao}, \citenamefont {Liu}, \citenamefont {Xiao}, \citenamefont {Han},
  \citenamefont {Lin}, \citenamefont {Wu}, \citenamefont {Lu}, \citenamefont
  {Hou}, \citenamefont {An}, \citenamefont {Wang}, \citenamefont {Cai},
  \citenamefont {Ho}, \citenamefont {He}, \citenamefont {Lortz}, \citenamefont
  {Zhang},\ and\ \citenamefont {Wang}}]{xu2017odd}%
  \BibitemOpen
  \bibfield  {author} {\bibinfo {author} {\bibfnamefont {S.}~\bibnamefont
  {Xu}}, \bibinfo {author} {\bibfnamefont {J.}~\bibnamefont {Shen}}, \bibinfo
  {author} {\bibfnamefont {G.}~\bibnamefont {Long}}, \bibinfo {author}
  {\bibfnamefont {Z.}~\bibnamefont {Wu}}, \bibinfo {author} {\bibfnamefont
  {Z.-q.}\ \bibnamefont {Bao}}, \bibinfo {author} {\bibfnamefont {C.-C.}\
  \bibnamefont {Liu}}, \bibinfo {author} {\bibfnamefont {X.}~\bibnamefont
  {Xiao}}, \bibinfo {author} {\bibfnamefont {T.}~\bibnamefont {Han}}, \bibinfo
  {author} {\bibfnamefont {J.}~\bibnamefont {Lin}}, \bibinfo {author}
  {\bibfnamefont {Y.}~\bibnamefont {Wu}}, \bibinfo {author} {\bibfnamefont
  {H.}~\bibnamefont {Lu}}, \bibinfo {author} {\bibfnamefont {J.}~\bibnamefont
  {Hou}}, \bibinfo {author} {\bibfnamefont {L.}~\bibnamefont {An}}, \bibinfo
  {author} {\bibfnamefont {Y.}~\bibnamefont {Wang}}, \bibinfo {author}
  {\bibfnamefont {Y.}~\bibnamefont {Cai}}, \bibinfo {author} {\bibfnamefont
  {K.~M.}\ \bibnamefont {Ho}}, \bibinfo {author} {\bibfnamefont
  {Y.}~\bibnamefont {He}}, \bibinfo {author} {\bibfnamefont {R.}~\bibnamefont
  {Lortz}}, \bibinfo {author} {\bibfnamefont {F.}~\bibnamefont {Zhang}}, \ and\
  \bibinfo {author} {\bibfnamefont {N.}~\bibnamefont {Wang}},\ }\href {\doibase
  10.1103/PhysRevLett.118.067702} {\bibfield  {journal} {\bibinfo  {journal}
  {Phys. Rev. Lett.}\ }\textbf {\bibinfo {volume} {118}},\ \bibinfo {pages}
  {067702} (\bibinfo {year} {2017})}\BibitemShut {NoStop}%
\bibitem [{\citenamefont {Fang}\ and\ \citenamefont
  {Stiles}(1968)}]{fang1968effects}%
  \BibitemOpen
  \bibfield  {author} {\bibinfo {author} {\bibfnamefont {F.~F.}\ \bibnamefont
  {Fang}}\ and\ \bibinfo {author} {\bibfnamefont {P.~J.}\ \bibnamefont
  {Stiles}},\ }\href@noop {} {\bibfield  {journal} {\bibinfo  {journal} {Phys.
  Rev.}\ }\textbf {\bibinfo {volume} {174}},\ \bibinfo {pages} {823} (\bibinfo
  {year} {1968})}\BibitemShut {NoStop}%
\bibitem [{\citenamefont {Nicholas}\ \emph {et~al.}(1988)\citenamefont
  {Nicholas}, \citenamefont {Haug}, \citenamefont {Klitzing},\ and\
  \citenamefont {Weimann}}]{Nicholas1988exchange}%
  \BibitemOpen
  \bibfield  {author} {\bibinfo {author} {\bibfnamefont {R.~J.}\ \bibnamefont
  {Nicholas}}, \bibinfo {author} {\bibfnamefont {R.~J.}\ \bibnamefont {Haug}},
  \bibinfo {author} {\bibfnamefont {K.~v.}\ \bibnamefont {Klitzing}}, \ and\
  \bibinfo {author} {\bibfnamefont {G.}~\bibnamefont {Weimann}},\ }\href@noop
  {} {\bibfield  {journal} {\bibinfo  {journal} {Phys. Rev. B}\ }\textbf
  {\bibinfo {volume} {37}},\ \bibinfo {pages} {1294} (\bibinfo {year}
  {1988})}\BibitemShut {NoStop}%
\bibitem [{\citenamefont {Kurganova}\ \emph {et~al.}(2011)\citenamefont
  {Kurganova}, \citenamefont {van Elferen}, \citenamefont {McCollam},
  \citenamefont {Ponomarenko}, \citenamefont {Novoselov}, \citenamefont
  {Veligura}, \citenamefont {van Wees}, \citenamefont {Maan},\ and\
  \citenamefont {Zeitler}}]{kurganova2011spin}%
  \BibitemOpen
  \bibfield  {author} {\bibinfo {author} {\bibfnamefont {E.}~\bibnamefont
  {Kurganova}}, \bibinfo {author} {\bibfnamefont {H.}~\bibnamefont {van
  Elferen}}, \bibinfo {author} {\bibfnamefont {A.}~\bibnamefont {McCollam}},
  \bibinfo {author} {\bibfnamefont {L.}~\bibnamefont {Ponomarenko}}, \bibinfo
  {author} {\bibfnamefont {K.}~\bibnamefont {Novoselov}}, \bibinfo {author}
  {\bibfnamefont {A.}~\bibnamefont {Veligura}}, \bibinfo {author}
  {\bibfnamefont {B.}~\bibnamefont {van Wees}}, \bibinfo {author}
  {\bibfnamefont {J.}~\bibnamefont {Maan}}, \ and\ \bibinfo {author}
  {\bibfnamefont {U.}~\bibnamefont {Zeitler}},\ }\href@noop {} {\bibfield
  {journal} {\bibinfo  {journal} {Physical Review B}\ }\textbf {\bibinfo
  {volume} {84}},\ \bibinfo {pages} {121407} (\bibinfo {year}
  {2011})}\BibitemShut {NoStop}%
\bibitem [{\citenamefont {Shashkin}\ \emph {et~al.}(2001)\citenamefont
  {Shashkin}, \citenamefont {Kravchenko}, \citenamefont {Dolgopolov},\ and\
  \citenamefont {Klapwijk}}]{shashkin2001indication}%
  \BibitemOpen
  \bibfield  {author} {\bibinfo {author} {\bibfnamefont {A.}~\bibnamefont
  {Shashkin}}, \bibinfo {author} {\bibfnamefont {S.}~\bibnamefont
  {Kravchenko}}, \bibinfo {author} {\bibfnamefont {V.}~\bibnamefont
  {Dolgopolov}}, \ and\ \bibinfo {author} {\bibfnamefont {T.}~\bibnamefont
  {Klapwijk}},\ }\href@noop {} {\bibfield  {journal} {\bibinfo  {journal}
  {Physical review letters}\ }\textbf {\bibinfo {volume} {87}},\ \bibinfo
  {pages} {086801} (\bibinfo {year} {2001})}\BibitemShut {NoStop}%
\bibitem [{\citenamefont {Zhu}\ \emph {et~al.}(2003)\citenamefont {Zhu},
  \citenamefont {Stormer}, \citenamefont {Pfeiffer}, \citenamefont {Baldwin},\
  and\ \citenamefont {West}}]{zhu2003spin}%
  \BibitemOpen
  \bibfield  {author} {\bibinfo {author} {\bibfnamefont {J.}~\bibnamefont
  {Zhu}}, \bibinfo {author} {\bibfnamefont {H.}~\bibnamefont {Stormer}},
  \bibinfo {author} {\bibfnamefont {L.}~\bibnamefont {Pfeiffer}}, \bibinfo
  {author} {\bibfnamefont {K.}~\bibnamefont {Baldwin}}, \ and\ \bibinfo
  {author} {\bibfnamefont {K.}~\bibnamefont {West}},\ }\href@noop {} {\bibfield
   {journal} {\bibinfo  {journal} {Physical review letters}\ }\textbf {\bibinfo
  {volume} {90}},\ \bibinfo {pages} {056805} (\bibinfo {year}
  {2003})}\BibitemShut {NoStop}%
\bibitem [{\citenamefont {Vakili}\ \emph {et~al.}(2004)\citenamefont {Vakili},
  \citenamefont {Shkolnikov}, \citenamefont {Tutuc}, \citenamefont
  {De~Poortere},\ and\ \citenamefont {Shayegan}}]{vakili2004spin}%
  \BibitemOpen
  \bibfield  {author} {\bibinfo {author} {\bibfnamefont {K.}~\bibnamefont
  {Vakili}}, \bibinfo {author} {\bibfnamefont {Y.}~\bibnamefont {Shkolnikov}},
  \bibinfo {author} {\bibfnamefont {E.}~\bibnamefont {Tutuc}}, \bibinfo
  {author} {\bibfnamefont {E.}~\bibnamefont {De~Poortere}}, \ and\ \bibinfo
  {author} {\bibfnamefont {M.}~\bibnamefont {Shayegan}},\ }\href@noop {}
  {\bibfield  {journal} {\bibinfo  {journal} {Physical review letters}\
  }\textbf {\bibinfo {volume} {92}},\ \bibinfo {pages} {226401} (\bibinfo
  {year} {2004})}\BibitemShut {NoStop}%
\bibitem [{\citenamefont {Gokmen}\ \emph {et~al.}(2007)\citenamefont {Gokmen},
  \citenamefont {Padmanabhan}, \citenamefont {Tutuc}, \citenamefont {Shayegan},
  \citenamefont {De~Palo}, \citenamefont {Moroni},\ and\ \citenamefont
  {Senatore}}]{gokmen2007spin}%
  \BibitemOpen
  \bibfield  {author} {\bibinfo {author} {\bibfnamefont {T.}~\bibnamefont
  {Gokmen}}, \bibinfo {author} {\bibfnamefont {M.}~\bibnamefont {Padmanabhan}},
  \bibinfo {author} {\bibfnamefont {E.}~\bibnamefont {Tutuc}}, \bibinfo
  {author} {\bibfnamefont {M.}~\bibnamefont {Shayegan}}, \bibinfo {author}
  {\bibfnamefont {S.}~\bibnamefont {De~Palo}}, \bibinfo {author} {\bibfnamefont
  {S.}~\bibnamefont {Moroni}}, \ and\ \bibinfo {author} {\bibfnamefont
  {G.}~\bibnamefont {Senatore}},\ }\href@noop {} {\bibfield  {journal}
  {\bibinfo  {journal} {Physical Review B}\ }\textbf {\bibinfo {volume} {76}},\
  \bibinfo {pages} {233301} (\bibinfo {year} {2007})}\BibitemShut {NoStop}%
\bibitem [{\citenamefont {Zhou}\ \emph {et~al.}(2017)\citenamefont {Zhou},
  \citenamefont {Lou}, \citenamefont {Zhang}, \citenamefont {Cheng},
  \citenamefont {Zhou},\ and\ \citenamefont {Chang}}]{zhou2017effective}%
  \BibitemOpen
  \bibfield  {author} {\bibinfo {author} {\bibfnamefont {X.}~\bibnamefont
  {Zhou}}, \bibinfo {author} {\bibfnamefont {W.-K.}\ \bibnamefont {Lou}},
  \bibinfo {author} {\bibfnamefont {D.}~\bibnamefont {Zhang}}, \bibinfo
  {author} {\bibfnamefont {F.}~\bibnamefont {Cheng}}, \bibinfo {author}
  {\bibfnamefont {G.}~\bibnamefont {Zhou}}, \ and\ \bibinfo {author}
  {\bibfnamefont {K.}~\bibnamefont {Chang}},\ }\href@noop {} {\bibfield
  {journal} {\bibinfo  {journal} {Physical Review B}\ }\textbf {\bibinfo
  {volume} {95}},\ \bibinfo {pages} {045408} (\bibinfo {year}
  {2017})}\BibitemShut {NoStop}%
\bibitem [{\citenamefont {Girvin}(2000)}]{girvin2000spin}%
  \BibitemOpen
  \bibfield  {author} {\bibinfo {author} {\bibfnamefont {S.~M.}\ \bibnamefont
  {Girvin}},\ }\href@noop {} {\bibfield  {journal} {\bibinfo  {journal}
  {Physics Today}\ }\textbf {\bibinfo {volume} {53}},\ \bibinfo {pages} {39}
  (\bibinfo {year} {2000})}\BibitemShut {NoStop}%
\bibitem [{\citenamefont {Falson}\ \emph {et~al.}(2015)\citenamefont {Falson},
  \citenamefont {Maryenko}, \citenamefont {Friess}, \citenamefont {Zhang},
  \citenamefont {Kozuka}, \citenamefont {Tsukazaki}, \citenamefont {Smet},\
  and\ \citenamefont {Kawasaki}}]{falson2015even}%
  \BibitemOpen
  \bibfield  {author} {\bibinfo {author} {\bibfnamefont {J.}~\bibnamefont
  {Falson}}, \bibinfo {author} {\bibfnamefont {D.}~\bibnamefont {Maryenko}},
  \bibinfo {author} {\bibfnamefont {B.}~\bibnamefont {Friess}}, \bibinfo
  {author} {\bibfnamefont {D.}~\bibnamefont {Zhang}}, \bibinfo {author}
  {\bibfnamefont {Y.}~\bibnamefont {Kozuka}}, \bibinfo {author} {\bibfnamefont
  {A.}~\bibnamefont {Tsukazaki}}, \bibinfo {author} {\bibfnamefont
  {J.}~\bibnamefont {Smet}}, \ and\ \bibinfo {author} {\bibfnamefont
  {M.}~\bibnamefont {Kawasaki}},\ }\href@noop {} {\bibfield  {journal}
  {\bibinfo  {journal} {Nature Physics}\ }\textbf {\bibinfo {volume} {11}},\
  \bibinfo {pages} {347} (\bibinfo {year} {2015})}\BibitemShut {NoStop}%
\bibitem [{\citenamefont {Movva}\ \emph {et~al.}(2017)\citenamefont {Movva},
  \citenamefont {Fallahazad}, \citenamefont {Kim}, \citenamefont {Larentis},
  \citenamefont {Taniguchi}, \citenamefont {Watanabe}, \citenamefont
  {Banerjee},\ and\ \citenamefont {Tutuc}}]{movva2017density}%
  \BibitemOpen
  \bibfield  {author} {\bibinfo {author} {\bibfnamefont {H.~C.}\ \bibnamefont
  {Movva}}, \bibinfo {author} {\bibfnamefont {B.}~\bibnamefont {Fallahazad}},
  \bibinfo {author} {\bibfnamefont {K.}~\bibnamefont {Kim}}, \bibinfo {author}
  {\bibfnamefont {S.}~\bibnamefont {Larentis}}, \bibinfo {author}
  {\bibfnamefont {T.}~\bibnamefont {Taniguchi}}, \bibinfo {author}
  {\bibfnamefont {K.}~\bibnamefont {Watanabe}}, \bibinfo {author}
  {\bibfnamefont {S.~K.}\ \bibnamefont {Banerjee}}, \ and\ \bibinfo {author}
  {\bibfnamefont {E.}~\bibnamefont {Tutuc}},\ }\href@noop {} {\bibfield
  {journal} {\bibinfo  {journal} {arXiv preprint arXiv:1702.05166}\ } (\bibinfo
  {year} {2017})}\BibitemShut {NoStop}%
\bibitem [{\citenamefont {Maryenko}\ \emph {et~al.}(2015)\citenamefont
  {Maryenko}, \citenamefont {Falson}, \citenamefont {Bahramy}, \citenamefont
  {Dmitriev}, \citenamefont {Kozuka}, \citenamefont {Tsukazaki},\ and\
  \citenamefont {Kawasaki}}]{maryenko2015spin}%
  \BibitemOpen
  \bibfield  {author} {\bibinfo {author} {\bibfnamefont {D.}~\bibnamefont
  {Maryenko}}, \bibinfo {author} {\bibfnamefont {J.}~\bibnamefont {Falson}},
  \bibinfo {author} {\bibfnamefont {M.}~\bibnamefont {Bahramy}}, \bibinfo
  {author} {\bibfnamefont {I.}~\bibnamefont {Dmitriev}}, \bibinfo {author}
  {\bibfnamefont {Y.}~\bibnamefont {Kozuka}}, \bibinfo {author} {\bibfnamefont
  {A.}~\bibnamefont {Tsukazaki}}, \ and\ \bibinfo {author} {\bibfnamefont
  {M.}~\bibnamefont {Kawasaki}},\ }\href@noop {} {\bibfield  {journal}
  {\bibinfo  {journal} {Physical review letters}\ }\textbf {\bibinfo {volume}
  {115}},\ \bibinfo {pages} {197601} (\bibinfo {year} {2015})}\BibitemShut
  {NoStop}%
\end{thebibliography}%

\end{document}